# Malware Classification using Deep Learning based Feature Extraction and Wrapper based Feature Selection Technique

Muhammad Furqan Rafique[1], Muhammad Ali[1], Aqsa Saeed Qureshi[1], Asifullah Khan[*,1,2,3], and Anwar Majid Mirza[4]

[1]Department of Computer Science, Pakistan Institute of Engineering & Applied Sciences, Nilore-45650, Islamabad, Pakistan
[2]PIEAS Artificial Intelligence Center, Pakistan Institute of Engineering & Applied Sciences, Nilore-45650, Islamabad Pakistan
[3]Centre for Mathematical Sciences, Pakistan Institute of Engineering and Applied Sciences, Nilore-45650, Islamabad, Pakistan
[4]College of Computer and Information Sciences, King Saud University, Riyadh, Saudi Arabia
asif@pieas.edu.pk

**Abstract**

In the case of malware analysis, categorization of malicious files is an essential part after malware detection. Numerous static and dynamic techniques have been reported so far for categorizing malware. This research presents a deep learning-based malware detection (DLMD) technique based on static methods for classifying different malware families. The proposed DLMD technique uses both the byte and ASM files for feature engineering, thus classifying malware families. First, features are extracted from byte files using two different Deep Convolutional Neural Networks (CNN). After that, essential and discriminative opcode features are selected using a wrapper-based mechanism, where Support Vector Machine (SVM) is used as a classifier. The idea is to construct a hybrid feature space by combining the different feature spaces to overcome the shortcoming of particular feature space and thus, reduce the chances of missing a malware. Finally, the hybrid feature space is used to train a Multilayer Perceptron, which classifies all nine different malware families. Experimental results show that proposed DLMD technique achieves log-loss of 0.09 for ten independent runs. Moreover, the proposed DLMD technique's performance is compared against different classifiers and shows its effectiveness in categorizing malware. The relevant code and database can be found at https://github.com/cyberhunters/Malware-Detection-Using-Machine-Learning.

## 1. Introduction

Malware can be considered any software that is intentionally designed to damage a computer, server, or any network [1]. Early day malware was not encrypted using complex cypher algorithms and thus were easily detected and classified by cross-matching some code pieces. However, with the recent polymorphism and metamorphism concepts like obfuscation, malware classification becomes a challenging and tedious task. Polymorphic malware exploits an encryption algorithm, which encrypts the code each time it iterates, while the encryption key remains constant, which makes it easier to detect [2]. In comparison, Metamorphic malware, which not only encrypts the code each time it iterates but, also change its encryption key, which makes it hard to detect [3]. It is observed that the total number of instances per day is growing drastically over the years, and thus manual inspection of malware analysis is considered ineffective. One of the main reasons for generating a high volume of malware samples is the extensive use of malware developers' obfuscation techniques, which means that malicious files from the same malware family (i.e. similar code and common origin) are modified continuously and obfuscated. Therefore, a generalized Machine Learning based malware analysis is considered a practical solution and can perform well on unseen samples. In this regard, static [4] and dynamic [5] analysis are used for malware detection and classification during training.

Static methods usually examine the malware's code (assembly or machine) without its execution [6]. Whereas, in dynamic methods, malware's behavior is monitored during its execution phase [7]. Both types of analysis have their own drawbacks. For example, in dynamic analysis, the susceptibility in the code cannot be dug out at the exact location, while static techniques do this job very well. On the other hand, the advantage of static analysis is that it can detect malware before its execution. Dynamic techniques allow to regain control of infected systems, which is not possible in static analysis.

In malware analysis, malware classification is important because categorizing various kinds of malware is important to know how they can contaminate personal computers, the risk level they pose, and how to defend them. In the case, malware is detected, it is assigned to the most appropriate malware family through a classification mechanism. There are numerous approaches for detecting malware in the wild; however, detecting a zero-day malware is still a challenging task.

In the literature, different malware classification techniques are reported. In Pascanu et al. approach, Recurrent Neural Network and Echo State Network are used as a feature extractor. Whereas extracted features are assigned to classifier to detect the malicious malware [8]. In Milosevic's work, clustering and classification based approach is used, whereby good results in terms of precision and recall rate are achieved [9]. To protect the computer user against automatically generated malware, Dahl et al. [10] proposed an interesting malware classification technique. In Dahl's technique, random projection technique is used to reduce the dimensionality of the input data. Finally, an ensemble of neural networks is trained, which reduced the error rate up to 0.42%. Kinable et al. [11] proposed a cal graph-based clustering technique to detect real malware instances. In their method, analyzers extract various characteristics from the program syntax and semantics such as operation codes and function call graph from the disassembled code, or string signatures and byte code n-grams from the hexa code, or different structural characteristics from the PE header, such as dependencies between APIs and DLLs[12]. Some other works [13] also explored the analysis of metadata such as the number of bitmaps, the size of import and export address tables beside the PE header's content. Cakir et al. [14] used only the assembly code and reported the log loss in the range of 0.06 to 0.03 on BIG 2015 by excluding the $5^{th}$ class. They used Word2vec [15] for feature representation and gradient boosting machine.

On the other hand, Microsoft research group [16] presented a neural network-based approach for malware analysis. During detection phase, function calls and call graph-based features are considered as important features, which resulted in detection accuracy of more than 90%. Cesare et al. [17] tried to solve the problem for malware encryption. Nowadays, most of the malware are using encryption or other obfuscation techniques. For malware, two different approaches are used to solve the encryption problem. The first approach is to decrypt or decode the malware binary. Second is to calculate entropy, which increases as compression ratio increases. This change in entropy can result in detection or classification of malware binary. Schultz et al. [18] used DLL function, system calls, and special characters as malware classification features. These features were extracted from byte file as strings.

Many static and dynamic analysis-based classification techniques are reported for the development of efficient malware classification system [19]–[28]. Amose et al. [19] evaluated the performance of different machine learning classifiers on android based dynamic malware detection. Similarly, Anderson et al. [20] also used dynamic analysis for malware detection. Anderson's approach is a graph-based malware detection technique, which is tested on different malware classification related problems. Petsas et al. [22] presented different analysis techniques. Petsas's technique information related to sensors, complexities related to the virtual machine, and static properties are used to detect android malware. Schmidt et al. [23] performed the static analysis by extracting function calls in the android environment. Then, lists of function calls and malware executables are compared, and then classification is performed.

Malware classification based on deep learning is also reported in the literature. Kalash et al. [25] used CNN for malware detection. In Kalash's technique, Maling and Microsoft dataset is used to check the effectiveness of their method. First, malware files are converted into image, whereby these grayscale images are provided as input to CNN. Their technique achieved an accuracy of 99.17 and 98.52 on Microsoft and Maling datasets respectively. Kims et al. [26] proposed an adversarial learning-based detection system to detect the zero-day malware attacks. In Kims's technique, fake malware is generated using Generative Adversarial Networks (GANs) [30]. Before the training of GAN, deep autoencoder is used to extract the generic features from data and then the knowledge is transferred to GAN to gain stability within the trained network. Similarly, another deep learning-based malware generation and detection scheme is reported by David et al. [27]. In David's technique, malware is the invariant and compact representation for malware is learned using deep belief network in combination with stacked denoising autoencoders.

In [28] work, a hybrid approach is used to detect different malware families within the BIG 2015 dataset. BIG 2015 dataset has become the benchmark for malware analysis, and different techniques are exploited on this dataset. Lee et al. [29] proposed instruction to the vector representation of assembly code and text-CNN for malware classification. Lee's approach can detect the intrusion without the execution of the virus and thus not harm the system. They created their dataset using Juliet.

Similarly, Temesguen Kebede et al. [31] use only byte file of viruses. They first convert them to images and train different autoencoders for the purpose of pretraining. Pretrained encoders are then concatenated with the softmax layer, and fine-tuning is performed. The reported accuracy was 99.15% on the BIG 2015 dataset, but they had excluded the $5^{th}$ malware family. In another malware detection approach, Ahmadi et al. [32] used both asm and byte files and reported an accuracy of 98.76% and log loss of 0.0094 on BIG 2015. Drew et al. [33] also used BIG 2015 and reported an accuracy of 98%. Their goal is to show how gene sequence classifier can be used for malware classification and how fast it works as compared to other hybrid techniques. Family classification is another important part of the BIG15 dataset because distinguishing between malware families is essential for the better understanding how they can infect computers and devices, the threat level they pose, and how to protect against them. Different machine learning techniques have been used so far for malware family classification. Some use opcodes or instructions of assembly code to predict representative classes, and some make images of machine language code to classify given malware, while others use hybrid approaches.

In the proposed deep learning-based malware detection (DLMD) technique, a hybrid dataset is first generated using deep learning-based feature extraction and wrapper-based feature selection technique. In the proposed DLMD technique, a information rich hybrid feature space is generated. The information extracted from multiple feature spaces brings diversity within the hybrid feature representation and used to classify different malware families effectively. BIG 2015 dataset is used for evaluating the performance of the proposed DLMD technique. Key features of our approach are:

- SVM is used to select informative feature space from the ASM file.
- CNN autoencoder is used for feature extraction from the byte file.
- Hybrid feature space is generated from both the extracted and selected feature spaces, which helps in overcoming the deficiencies associated with different feature spaces.
- The final classification of malware families is performed using an MLP.

## 2. Proposed DLMD technique for malware classification

The proposed DLMD technique detects different malware families provided in BIG15 malware detection related dataset. The dataset contains both byte and asm files against nine different malware families. Therefore, the proposed DLMD methodology used both byte and asm files of the dataset for feature engineering. First, two deep CNNs are used to extract features from byte file. Whereas, wrapper based feature selection technique is applied on asm file for the purpose of selecting informative features[34]. After that, hybrid dataset is created by combining all the selected and extracted features. In the end, hybrid feature space is used to train the Multilayer perceptron, which finally classifies the nine different malware families. Block diagram of the proposed DLMD technique is shown in Figure 1.

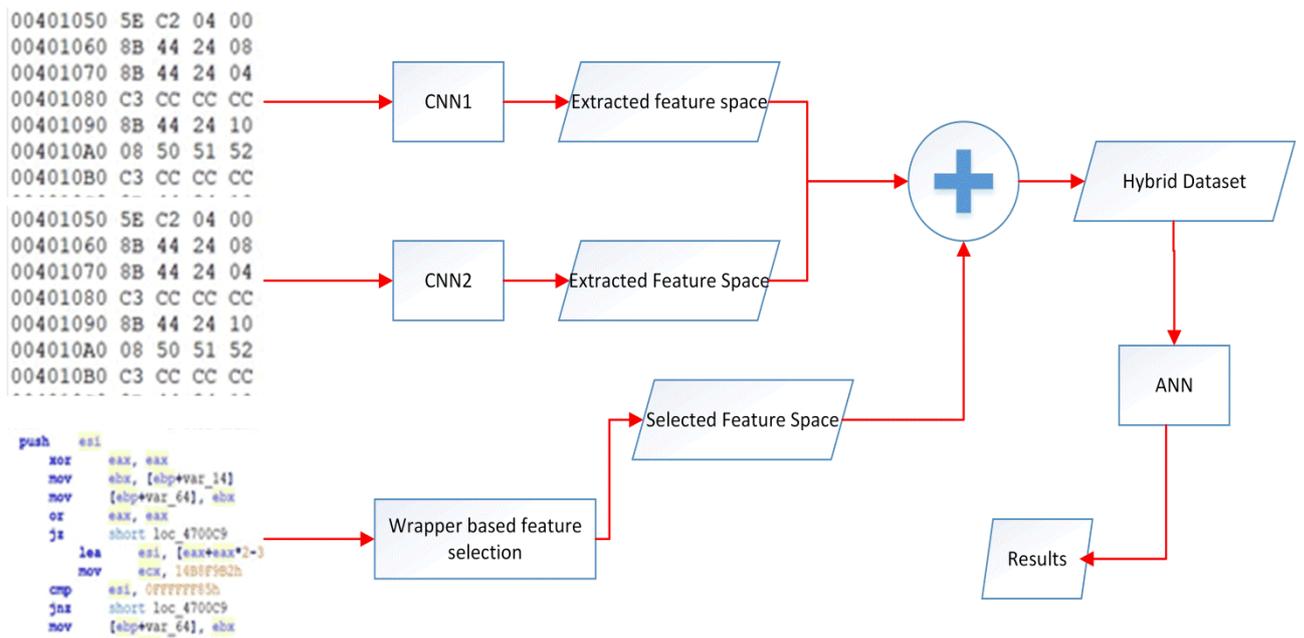

Figure 1 Proposed DLMD technique

### 2.1. Background related to Deep learning model used in DLMD technique

In the proposed DLMD technique, different deep learning models are used to improve the input feature representation. Details related to these models are discussed in below-mentioned subsections.

#### 2.1.1. CNN

CNN a special type of ANN, which works on the principle of local connectivity. Hence, they can exploit local correlation by ensuring local connectivity between neurons of adjacent layers. Convolutional layers are the building block of CNN. During training, kernel having fixed height and width is convolved through the whole matrix of input and output matrix is generated. During training, kernels are evolved automatically through backpropagation algorithm. Hence, the output is the results of the convolved kernel at each position in the input matrix. Following are the major steps of any generic CNN architecture[35][36]:

1) Convolve learnable filters on the input matrix
2) Apply pooling and activation on the resulted values
3) Apply standard ANN to the features resulted by step 2

Pooling is another important layer within CNN architecture. Pooling layers within CNN introduces nonlinear downsampling. Different versions of pooling layer include min, max, and average. Maximum pooling is used widely due to its reduced computation and transition invariance property. After stacking of multiple convolutional and pooling layers, the fully connected layer is used at the end. This layer maps the features that are extracted from the multiple convolutions and pooling layers to the output. This also adds the multilayer perceptron's functionality in the CNN, but it does not work directly on the input, but on the extracted features by applying kernels on the input.

### 2.1.2. Autoencoder

Autoencoder is an ANN used to learn a representation of data. It is mostly used for dimensionality reduction. They can learn low-level representations of the data and their projection back to the original data. Three main components of autoencoder are:

1. Encoding Architecture that is used to reduce the dimensionality of input data mostly in a series of layers with nodes decreasing.
2. Low-Level representation of data after it is passed through the encoding architecture.
3. Decoding architecture that has a series of layers with a mostly increasing number of layers that results in the projection of data back to original form from low-level representation.

In order to bring generalization during the training of autoencoder, L1 and L2 weight regularization terms are added within the loss function. Regularization is a basic principle of machine learning. Its purpose is to stop the model from being overfitted. It is applied in the minimization of the cost function. In L1 regularization sum of weights are added to the cost so that they are to be minimized. Hence, it produces very sparse matrices containing many zeros. Whereas, in L2 regularization sum of the square of weights are added to the cost function. L2 regularization is also used in the cost function to avoid overfitting.

### 2.2. Dataset

The dataset used for the evaluation of the proposed DLMD technique is BIG 2015, published by Microsoft on the Kaggle platform. This data has become a benchmark for researchers and cited in more than 50 papers[28]. The dataset is comprised of 0. 5 terabytes, containing 10868 samples for training and 10873 for testing. Up till now, labels are not provided against the test sample, so only train data is used for this research. The dataset consists of 9 malware families, whereas the frequency distribution of different families is shown graphically and in tabular form in Figure 2 and Table 1, respectively. The figure shows a high imbalance between classes in the malware dataset. The most abundant class is Kelihos_ver3 with approximately five thousand samples, where Simda class has the least samples of nearly forty.

Description related to different malware families is discussed below:

- **Ramnit:** It steals user credentials such as password credit card information and halts security software.
- **Lollipop:** It is an adware that shows ads on the browser; it also allows a hacker to monitor user traffic.
- **Kelihos_ver1 and Kelihos_ver3:** Trojan types can fully control user pc and spread by sending spam email from user pc to others.
- **Vundo:** It could be responsible for pop-up ads and installing other malicious content.
- **Simda:** These type of malware snatch the passwords from user pc and create a backdoor for hackers.
- **Traceur:** Using this malware attack, Author generates revenue by showing bogus advisement on search engines.
- **Obfuscator.ACY:** These are considered as obfuscated malware, and their purpose could be any of the malware mentioned above.
- **Gatak:** This is also a type of Trojan that seems legitimate but infects a computer with its malicious code.

In BIG 2015 training dataset, against each malware, two files are provided one is byte file and the other one is assembly code (.asm file). Description of both files is discussed below.

**Table 1 Number of samples in BIG15 dataset**

| Class | Frequency |
|---|---|
| Ramnit | 1541 |
| Lollipop | 2478 |
| Kelihos_ver3 | 2942 |
| Vundo | 475 |
| Simda | 42 |
| Tracur | 751 |
| Kelihos_ver1 | 398 |
| Obfuscator.ACY | 1228 |
| Gatak | 1013 |

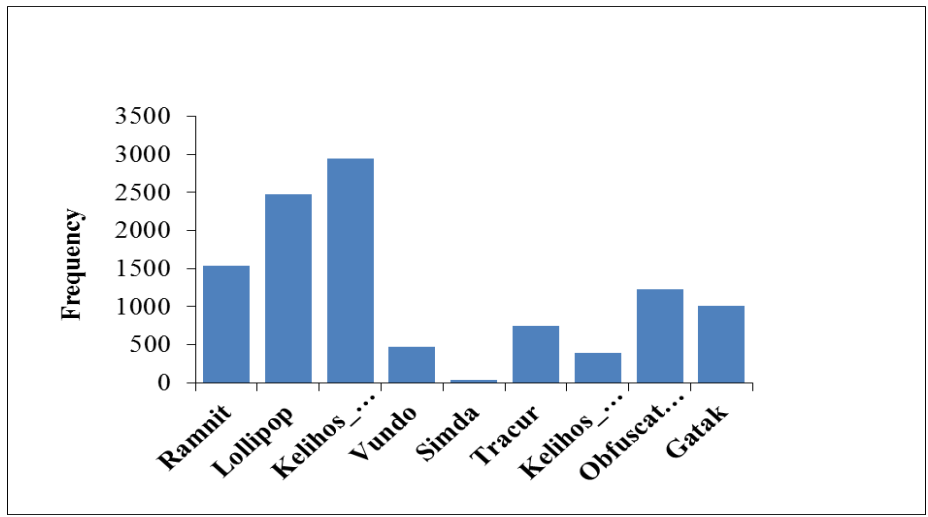

**Figure 2 Frequency of malware classes**

### 2.2.1. ASM file

Asm files are basically the malware assembly code that contains the information related to the function calls and variable allocation. A sample of the ASM file is shown in Figure 3. Major sections of an ASM file are comprised of bss and text section. However, other sections having information related to secondary functions may also be present. Different ASM files usually contain a different number of sections. These portions' existence depends upon the type of malware, i.e. obfuscated virus sections will be different from ramnit malware. Some of the basic sections and their purpose are mentioned below:

- .text: This section contains the actual code of the malware
- .data: This section contains the initialized variables or data
- .rdata: Read-only data or constants that are not allowed to change during the running of the program are placed here.
- .bss: This section contains uninitialized variables, that can be used during execution
- .idata: It contains the information about the directories or the program that code will import during execution
- .edata: It contains information about the data that malware will export during execution.
- .rsrc: It contains actual resources of the program.

```
.text:00402078 50                              push    eax
.text:00402079 8D 44 24 20                     lea     eax, [esp+2Ch+var_C]
.text:0040207D 64 A3 00 00 00 00               mov     large fs:0, eax
.text:00402083 33 C0                           xor     eax, eax
.text:00402085 6A 05                           push    5               ; MaxCount
.text:00402087 68 40 BB 42 00                  push    offset a9gw0p   ; "9gw0p"
.text:0040208C 8D 4C 24 0C                     lea     ecx, [esp+34h+var_28]
.text:00402090 89 44 24 30                     mov     [esp+34h+var_4], eax
.text:00402094 C7 44 24 24 0F 00 00 00         mov     [esp+34h+var_10], 0Fh
.text:0040209C 89 44 24 20                     mov     [esp+34h+var_14], eax
.text:004020A0 88 44 24 10                     mov     byte ptr [esp+34h+var_24], al
.text:004020A4 E8 B5 C6 03 00                  call    sub_43E75E
.text:004020A9 83 7C 24 1C 10                  cmp     [esp+2Ch+var_10], 10h
.text:004020AE 72 0D                           jb      short loc_4020BD
.text:004020B0 8B 44 24 08                     mov     eax, [esp+2Ch+var_24]
```

**Figure 3 Assembly code of ASM file**

Each section has its own type of opcodes. As the proposed technique only deals with opcodes, so redundant information is removed from asm files. Further, for each malware sample, the term frequency against unique opcodes are recorded, as shown below:

| ID | LABELS | push | mov | sub | lea | call | pop |
|---|---|---|---|---|---|---|---|
| 01kcPWA9K2BOxQeS5R | 1 | 81 | 89 | 5 | 36 | 53 | 19 |
| 04EjldbPV5e1XroFOpiN | 1 | 5927 | 9764 | 311 | 1230 | 2900 | 1527 |
| 05EeG39MTRrI6VY21DP | 1 | 915 | 2415 | 73 | 157 | 461 | 376 |
| 05rJTUWYAKNegBk2wE | 1 | 28620 | 32566 | 1240 | 5819 | 10686 | 6384 |
| 0AnoOZDNbPXIr2MRBS | 1 | 781 | 2624 | 221 | 344 | 462 | 539 |
| 0AwWs42SUQ19mI7eD( | 1 | 3623 | 5174 | 175 | 1139 | 1957 | 1370 |

### 2.2.2. Byte file

Byte files are the hexadecimal representation of the portable executable (PE) of the malware. Each row of the file is called a record. In the given dataset, it has two portions. The first one is the offset of the memory address of the instruction. The second portion is the instruction itself represented in the pair of hexa decimals. The IDA disassembler tool can convert any portable executable file into byte file. These byte files are the direct mapping of assembly instructions that are present in the asm file.

Byte files converted into images, as shown in figure 4. Each hexadecimal pair is treated as a single decimal number which serves as a pixel value of the image. These images than resized to a standard dimension of 32x32.

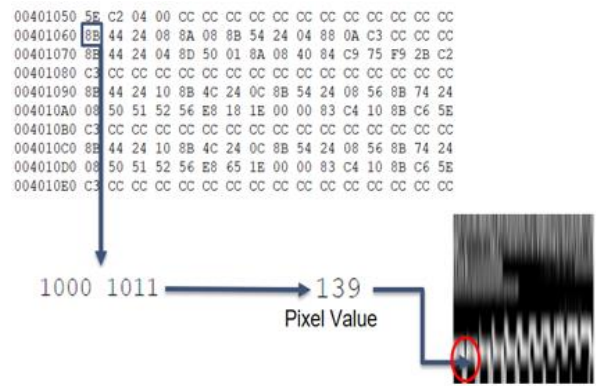

**Figure 4 Byte File to image conversion**

### 2.2.3. Normalization

Min-max normalization ($X_{norm} = \dfrac{X - X_{min}}{X_{max} - X_{min}}$) is applied on both images and term frequency table.

### 2.2.4. Division of Dataset

To check the performance of the proposed DLMD technique, the dataset is divided in a stratified manner across training, validation and test set. Out of total data $(D)$, 25% is reserved as test data $(D_{TEST})$. Whereas, out of the remaining 75% of $D$ $(D_T)$, 25% of $D_T$ $(D_{VAL})$ is reserved as validation data. Similarly, 75% $D_T$ $(D_{TRAIN})$ is used during the training of the proposed DLMD technique. During training, the proposed DLMD technique is trained on the training set $(D_{TRAIN})$, and parameters are optimized based on the trained network's performance on validation data. After selecting the parameters of the proposed DLMD technique, performance is checked on test data ($(D_{TEST})$). Division of BIG 15 dataset is shown in Figure 5.

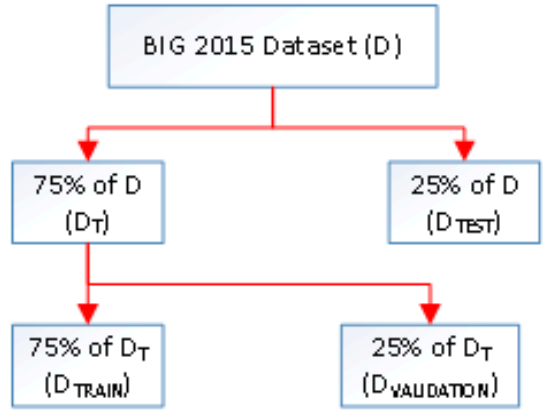

**Figure 4 Distribution of data**

### 2.3. Hybrid feature representation using deep learning and Wrapper based feature selection

In the proposed DLMD technique, the hybrid dataset is generated by concatenating the feature space extracted by passing original feature space through two different CNN models and feature set selected using the wrapper-based feature selection technique. At the end, hybrid feature space is provided as input to MLP. The details related to selection and extraction techniques are discussed in below-mentioned sections.

### 2.3.1. Feature Extraction using CNN

In the proposed DMDL technique, feature extraction is performed by using the CNN architectures that trained in two different ways. For one type of feature extraction, convolutional layers of CNN are used to learn the distinct representation of malware to improve the proposed DMDL classification model's performance. Therefore, in this regard, five layers deep CNN was trained from scratch on image representation (byte file) of 9 different malware classes during the training phase. The best model state was saved on which log loss is minimized on the validation data set. The train, validation, and test sets are passed separately through that saved model to get nine probabilities (nine extracted features) against each malware. The architecture of CNN that is designed to extract the feature space using the five-layered CNN model is shown in Figure 6

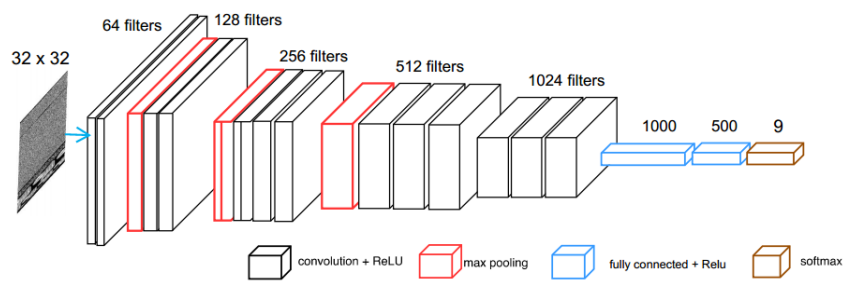

Figure 5 Five layered CNN architecture

The second type of feature extraction is also performed by using CNN, in such a way, convolutional layers of CNN are trained using convolutional autoencoders. In this regard, initially, two convolutional autoencoders (CAE) are trained on original byte files. Whereas, input to the second autoencoder is the original encoded data by the first encoder. In a second phase, CAE's pre-trained encoding layers are used as convolution layers of a CNN with two fully connected layers for fine-tuning purpose, as explained in Figure 7. During the fine-tuning best model state is saved on which log loss is minimum for validation data. After training, all the dataset (train, validation, and test) passes through the best model to get nine probabilities against each malware sample as features.

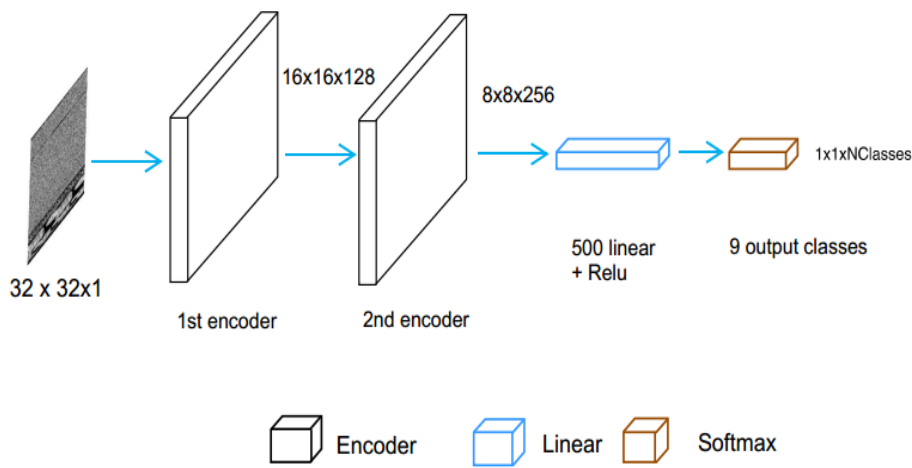

Figure 6 Architecture of Feature extraction with pretrained layers

### 2.3.2. Wrapper based Feature selection

In proposed DLMD methodology, features are selected using wrapper-based techniques in which SVM with RBF kernel is used as a classifier for selecting important opcodes from ASM file. For feature selection, both forward and backward selection approaches are used. It starts with ten features and does increment of 10 (forward selection). After optimization of each SVM, it is noted that once features reach 120, SVM performance started to degrade, so decrement of 1 (backward selection) is done for better results. In the proposed DLMD technique, 116 features are selected using the wrapper-based feature selection technique. Feature selection using the wrapper-based technique is shown in Figure 8.

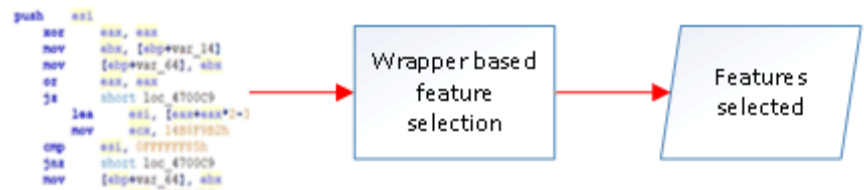

**Figure 7 Wrapper based feature selection technique**

### 2.4. Classification of malware families on the hybrid dataset using MLP

After feature selection and extraction, the hybrid dataset is generated by concatenating nine features extracted through two different CNN based feature extractors (total 18 features) with 116 selected opcodes (selected using the wrapper-based technique). After that parameter of a multi perceptron is optimized on the normalized hybrid dataset

## 3. Implementation Details

All experimental work and simulations related to the proposed work were performed on the Desktop computer with 61.1 GiB memory, Intel Core i7-6700 CPU@ 3.40GHzx8, and GeForce GTX 1070/PCIe/SSE2 and disk memory 919.9 GB; The Operating system used was Ubuntu 16.04 LTS. The major framework used for the project is Anaconda and for deep learning is Pytorch.

### 3.1. Parameter setting of proposed DLMD technique

In the proposed DLMD technique, two different types of CNN are used as feature extractors. Therefore, in this regard, a five-layer deep CNN was trained from scratch on image representation (byte file) of 9 different malware classes. During the training phase, the best model state was saved on which log loss is minimum on the validation dataset. The parameters that are optimized during training are shown in Table 2. Whereas, details related to the architecture of CNN are discussed in Table 3.

The second type of feature extractor used in the proposed DLMD work is CNN, whose layers are based on CAE. The parameters that are selected during optimization are shown in Table 4. The architecture of both autoencoders and CNN with pre-trained CAE layers are shown below in Table 5, Table 6, and Table 7.

Table 2 Parameters of CNN

| Parameters | Value |
|---|---|
| layers | 5 |
| epochs | 30 |
| Batch size | 20 |
| Learning rate | 0.001 |

Table 3 Architecture of CNN

| Layers | Kernal_size | Filters | Input | Output |
|---|---|---|---|---|
| 1st 2D conv | 3x3 | 64 | 32x32x1 | 32x32x64 |
| LeakyRelu | | | 32x32x64 | 32x32x64 |
| Maxpool2D | 2x2 | | 32x32x64 | 16x16x64 |
| Batch_norm | | | 16x16x64 | 16x16x64 |
| 2nd 2D conv | 3x3 | 128 | 16x16x64 | 16x16x128 |
| LeakyRelu | | | 16x16x128 | 16x16x128 |
| Maxpool2D | 2x2 | | 16x16x128 | 8x8x128 |
| Batch_norm | | | 8x8x128 | 8x8x128 |
| 3rd 2D conv | 3x3 | 256 | 8x8x128 | 8x8x256 |
| LeakyRelu | | | 8x8x256 | 8x8x256 |
| Maxpool2D | 2x2 | | 8x8x256 | 4x4x256 |
| Batch_norm | | | 4x4x256 | 4x4x256 |
| 4th 2D conv | 3x3 | 512 | 4x4x256 | 4x4x512 |
| LeakyRelu | | | 4x4x512 | 4x4x512 |
| Maxpool2D | 2x2 | | 4x4x512 | 2x2x512 |
| Batch_norm | | | 2x2x512 | 2x2x512 |
| 5TH 2D conv | 3x3 | 1024 | 2x2x512 | 2x2x1024 |
| LeakyRelu | | | 2x2x1024 | 2x2x1024 |
| Batch_norm | | | 2x2x1024 | 2x2x1024 |

| | | | | | |
|---|---|---|---|---|---|
| 1st Linear layer | | | | **2x2x1024** | **1000** |
| LeakyRelu | | | | 1000 | 1000 |
| 2nd Linear layer | | | | **1000** | **500** |
| LeakyRelu | | | | 500 | 500 |
| 3rd Linear layer | | | | **500** | **9** |
| Softmax | | | | 9 | 9 |

**Table 4 Parameters of convolutional autoencoder**

| Parameters | Value |
|---|---|
| Weight decay | 0.00001 |
| epochs | 100 |
| Batch size | 20 |
| Learning rate | 0.001 |

**Table 5 Architecture of 1st autoencoder**

| | Layers | Kernal_size | Filters | Stride | Padding | Input | Output |
|---|---|---|---|---|---|---|---|
| Encoding layer | **2D conv** | **3x3** | **128** | **1** | **1** | **32x32x1** | **32x32x128** |
| | Relu | | | | | 32x32x128 | 32x32x128 |
| | Maxpool2D | 2x2 | | 2 | | 32x32x128 | 16x16x128 |
| Decoding layer | **ConvTranpose** | **2x2** | **1** | **2** | | **16x16x128** | **32x32x1** |
| | Relu | | | | | 32x32x1 | 32x32x1 |

**Table 6 Architecture of 2nd CNN**

| | Layers | Kernal_size | Filters | Stride | Padding | Input | Output |
|---|---|---|---|---|---|---|---|
| Encoding layer | **2D conv** | **3x3** | **128** | **1** | **1** | **16x16x128** | **16x16x256** |
| | Relu | | | | | 16x16x256 | 16x16x256 |
| | Maxpool2D | 2x2 | | 2 | | 16x16x256 | 8x8x256 |
| Decoding layer | **ConvTranpose** | **2x2** | **128** | **2** | | **8x8x256** | **16x16x128** |
| | Relu | | | | | 16x16x128 | 16x16x128 |

**Table 7 Architecture of CNN with pre-trained layers**

| Layers | Input | Output |
|---|---|---|
| 1st encoder layer | 32x32x1 | 16x16x128 |
| 2nd encoder layer | 16x16x128 | 8x8x256 |
| Linear layer | 8x8x256 | 500 |
| Relu | 500 | 500 |
| Linear layer | 500 | 9 |
| Softmax | 9 | 9 |

## 4. Performance Evaluation Measures

In the proposed DLMD technique, log loss is used as an evaluation measure. Log loss is the cross-entropy between correct labels and predicted labels. The exact formula for the log loss evaluation has been shown in equation 1.

$$Logloss = -\frac{1}{N}\sum_{i=1}^{N}\sum_{j=1}^{M} y_{ij} \log(\rho_{ij}) \tag{1}$$

Here, N is the number of samples, and M is the number of classes. y represents the true label of the class, and $\rho$ is the probability of the given sample. A log that is mentioned in the above formula is Natural Logarithm.

Accuracy is another evaluation measure used for evaluating the performance of the proposed DLMD technique. Mathematically accuracy can be defined as below:

$$\frac{TP+TN}{TP+FP+TN+FN} \tag{2}$$

In the above equation, TP and TN are the number of positive and negative samples, respectively which are correctly classified by the classifiers. Whereas, FP and FN are the number of positive and negative samples, respectively, which are misclassified by the classifiers.

## 5. Results

In the proposed DLMD technique, SVM is used as a feature selector and simple CNN and CNN autoencoder are used as a feature extractor. Afterwards, MLP is used as a classifier.

## 5.1. Comparative Evaluation of the Baseline models on Image dataset on Test Set Using Ten Independent Runs.

Before checking the performance of the proposed DLMD technique, performance is checked on simple baseline classifiers. Table 8 shows mean log loss score for ten independent runs for SVM, simple multi-layer perceptron and CNN along with the optimised parameters during training. Table 8. depicts that CNN is performing better with less standard deviation, which means that image like features better represents that feature space of malware. Performance obtained through baseline classifiers shows that better results can be achieved from CNN due to its nature of learning localized image structure.

Table 8 Baseline classifiers

| Model | Parameters | Mean log loss | Std |
|---|---|---|---|
| SVM (Linear) | C=1 | 0.8948 | 0.0122 |
| SVM (RBF) | C=10, gamma=0.1 | 0.1882 | 0.0069 |
| MLP | 3 Layer MLP (1000,500,100) | 0.2828 | 0.0087 |
| CNN | Conv 3, Pool 3, Linear 2 | 0.1514 | 0.006 |

## 5.2. Performance of Proposed DLMD technique

Secondly, we applied our proposed DLMD technique on the hybrid dataset. This resulted in less class biasness, as shown in the confusion matrix and decreased log loss. SVM and CNN resulted in better feature selector and extractor, respectively. Table 9 shows the proposed DLMD technique's performance in terms of Log loss and accuracy against ten independent runs.

Table 9 Performance of proposed DLMD technique

| Sr. | Log loss | Accuracy |
|---|---|---|
| 1 | 0.0951 | 97.6014 |
| 2 | 0.0964 | 97.4907 |
| 3 | 0.0951 | 97.6014 |
| 4 | 0.0972 | 97.4538 |
| 5 | 0.0959 | 97.5645 |
| 6 | 0.0954 | 97.6383 |
| 7 | 0.0971 | 97.4538 |
| 8 | 0.0957 | 97.5645 |
| 9 | 0.0954 | 97.5276 |
| 10 | 0.0962 | 97.6383 |
|  | 0.0961±0.0008 | 97.5535±0.0008 |

Confusion matrix for the above-applied approach is as below. It can be clearly seen that it is not biased toward most of the classes as seen in the case of SVM, where most of the misclassified samples were assigned to first class.

The experiment result shows that our proposed approach achieved lower log loss with a good confusion matrix compared with other baseline classifier and selected feature space is more effective and helpful for classification of malware families

**Table 10 Confusion matrix**

|    | C1  | C2  | C3  | C4  | C5 | C6  | C7 | C8  | C9  |
|----|-----|-----|-----|-----|----|-----|----|-----|-----|
| C1 | **367** | 1   | 0   | 3   | 0  | 4   | 0  | 7   | 1   |
| C2 | 0   | **610** | 2   | 1   | 0  | 2   | 1  | 3   | 0   |
| C3 | 0   | 1   | **734** | 0   | 0  | 0   | 0  | 0   | 0   |
| C4 | 0   | 0   | 0   | **116** | 0  | 1   | 0  | 1   | 0   |
| C5 | 0   | 1   | 0   | 0   | **9** | 0 | 0  | 0   | 0   |
| C6 | 2   | 2   | 0   | 0   | 1  | **174** | 0 | 6 | 2   |
| C7 | 0   | 0   | 1   | 0   | 0  | 0   | **97** | 0 | 1   |
| C8 | 9   | 5   | 0   | 3   | 0  | 2   | 0  | **284** | 3 |
| C9 | 0   | 1   | 0   | 0   | 0  | 0   | 1  | 1   | **250** |

**5.3. Comparison of the Proposed method with Existing Techniques and our baseline models**

During the experiment, 25% of total data is reserved as test data for checking the proposed DLMD technique with commonly used baseline classifiers. However, for fair comparison of the proposed DLMD technique with Drew's technique [33] 10% of total data is used as test data, as this proportion is defined in the Drew's work.

Table 11 shows the performance comparison of proposed DLMD (with different test data ratios) with SVM, MLP, CNN, and Drew's technique in terms of mean log loss against ten independent runs. Results clearly show that the proposed DLMD technique is good in terms of mean log loss; also standard deviation against ten independent runs is very low in comparison to other classifiers. The low value of log loss shows that the proposed technique is stable and reliable. As the proposed technique is an ensemble-based hybrid approach that automatically generated feature space through the different architecture of CNN along with informative features selected through the wrapper-based feature selection technique. Hybrid feature space (which contains diverse information ) improves the proposed DLMD technique's performance compared to when only original feature space is provided as input to multiple classifiers.

Figure 9 compares the proposed DLMD technique with other commonly used classifiers and Drew's technique [33]. This figure clearly depicts that the hybrid feature space generated after feature selection and extraction phase helps improve the performance of the overall malware classification system. Also, the mean log loss of the proposed DLMD technique is low when 10% of data is reserved as test data compared to 25% of data reserved as test data. Comparison of the proposed technique with Drew's technique shows that the proposed technique is more stable and reliable.

**Table 11 Performance comparison of proposed DLMD technique with other techniques**

| Model | Mean log loss |
|---|---|
| Linear-SVM(25%test data) | 0.8948±0.0122 |
| RBF-SVM (25%test data) | 0.1882±0.0069 |
| MLP(25% test data) | 0.2828±0.0087 |
| CNN(25% test data) | 0.1514±0.006 |
| DLMD(25% test data) | 0.0961±0.0008 |
| DLMD(10% test data) | 0.0378±0.0007 |
| Drew's technique [33] (10% test data) | 0.0479 |

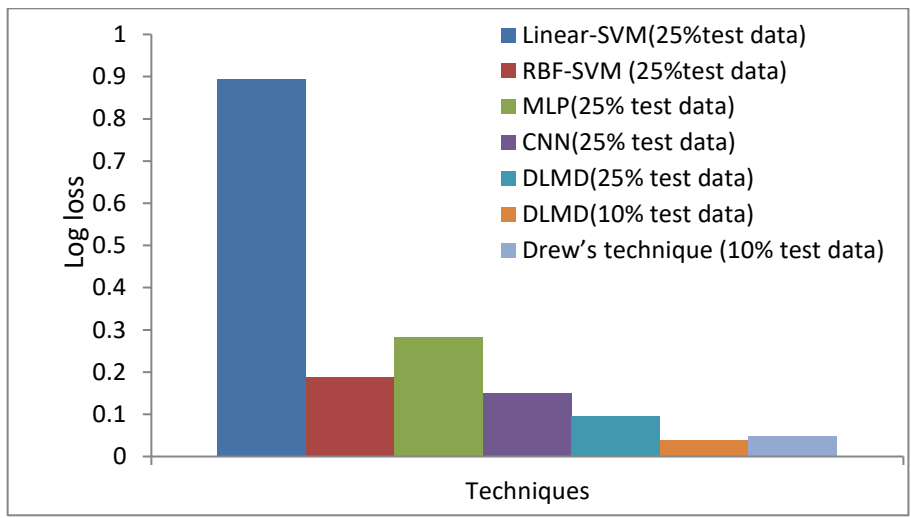

**Figure 8 Comparison of proposed DLMD technique with other techniques**

## 6. Conclusion

In the proposed DLMD technique, the byte and ASM file are utilized to develop a hybrid dataset for developing effective malware classification system. ASM files were used to obtain the count of the opcodes or words present in the assemble source file. This approach is platform dependent but uses simple features as opcode count that are easy to extract. Deep CNN is used as a feature extractor whereas wrapper-based feature selection technique is used as a feature selector for forming a hybrid dataset. In the end, the performance of hybrid dataset is evaluated on commonly used classifiers. This reduced loss is that extracted and selected feature space from byte and ASM files provides diversity. The shortcomings of extracted feature space are overcome by selected feature space, which resulted in better feature representation for ANN to classify the malware. In the future, the idea of hybrid feature space generated using the proposed DLMD technique will help researchers develop an efficient Malware detection system capable of detecting new attacks. The proposed DLMD technique is compared against different existing techniques and shows its effectiveness in categorizing malware.

## ACKNOWLEDGMENT

We acknowledge Pakistan Institute of Engineering and Applied Sciences for the healthy research environment which led to research work presented in this paper.

**Conflict of Interest**

The authors declare that they have no conflict of interest.